\documentclass[12pt]{article}
\parskip=\medskipamount                 % space between paragraphs
\thicklines                     % thick straight lines for pictures
\usepackage{amsfonts,amssymb,amsmath,amscd}
\usepackage{bbm}
\usepackage[ps,dvips,matrix,arrow,frame,import,curve,color]{xy}
\newcommand{\raa}{\rightarrow}

\newcommand{\Tr}{{\rm Tr}}

\newcommand{\ZZ}{{\mathbb Z}}
\newcommand{\RR}{{\mathbb R}}
\newcommand{\CC}{{\mathbb C}}

\newcommand{\cS}{{\mathcal S}}

\newcommand{\cO}{{\mathcal O}}
\newcommand{\bpartial}{{\bar\partial}}
\newcommand{\eps}{\epsilon}

\newcommand{\HK}{{hyper-K\"ahler}}

\newcommand{\Hom}{{\rm Hom}}

\newcommand{\cV}{{\mathcal V}}

\newcommand{\cE}{{\mathcal E}}

\newcommand{\cN}{{\mathcal N}}
\newcommand{\End}{{\rm End}}

\newcommand{\STr}{{\rm STr}}

\newcommand{\bI}{{\bar I}}
\newcommand{\bJ}{{\bar J}}

\newcommand{\cF}{{\mathcal F}}

\newcommand{\frg}{{\mathfrak g}}
\newcommand{\sE}{{\mathcal E}}
\newcommand{\hQ}{{\hat Q}}
\newcommand{\frG}{{\mathfrak G}}
\newcommand{\vol}{{\rm vol}}

\newcommand{\gl}{{\mathfrak {gl}}}
\newcommand{\osp}{{\mathfrak {osp}}}
\newcommand{\PP}{{\mathbb P}}
\renewcommand{\sl}{{\mathfrak {sl}}}
\newcommand{\hP}{{\hat P}}
\newcommand{\wV}{{\mathcal W}}
\newcommand{\sfW}{{\mathsf W}}

\def\be{\begin{equation}}
\def\ee{\end{equation}}
\def\bear{\begin{eqnarray}}
\def\eear{\end{eqnarray}}
\def\nn{\nonumber}

\def\half{{ \frac{1}{2} }}

\def\wdg{{\wedge}}

\def\Tau{{\cal T}}
\def\del{{\partial}}
\def\a{{\alpha}}
\def\b{\beta}

\def\eps{{\epsilon}}

\def\cL{{{\mathcal L}}}

\def\p{{\partial}}
\def\delbar{{\overline {\partial}}}

\def\wdg{{\wedge}}
\def\vert{{|}}

\def\xId{ \mathbbm{1} }

\begin{document}

\title{Chern-Simons-Rozansky-Witten topological field theory}

\author{Anton Kapustin and Natalia Saulina \\ {\it \small California
Institute of Technology}}

\begin{titlepage}
\maketitle

\abstract{We construct and study a new topological field theory in
three dimensions. It is a hybrid between Chern-Simons and
Rozansky-Witten theory and can be regarded as a
topologically-twisted version of the $N=4$ $d=3$ supersymmetric
gauge theory recently discovered by Gaiotto and Witten. The model
depends on a gauge group $G$ and a \HK\ manifold $X$ with a
tri-holomorphic action of $G$. In the case when $X$ is an affine
space, we show that the model is equivalent to Chern-Simons theory
whose gauge group is a supergroup. This explains the role of Lie
superalgebras in the construction of Gaiotto and Witten. For general
$X$, our model appears to be new. We describe some of its
properties, focusing on the case when $G$ is simple and $X$ is the
cotangent bundle of the flag variety of $G$. In particular, we show
that Wilson loops are labeled by objects of a certain category which
is a quantum deformation of the equivariant derived category of
coherent sheaves on $X$.}

\end{titlepage}

\section{Introduction and summary}

Recently, an interesting new class of three-dimensional gauge
theories with $N=4$ supersymmetry has been discovered \cite{GW}. The
distinguishing feature of these field theories is that the gauge
field kinetic term is the Chern-Simons term, while the usual
Yang-Mills kinetic term is absent.

The theories constructed in \cite{GW} contain a Chern-Simons gauge
field interacting with $N=4$ hypermultiplets. In the limit of
vanishing gauge coupling hypermultiplets are described by an $N=4$
$d=3$ sigma-model whose target $X$ is required to be \HK\ by
supersymmetry. For nonvanishing gauge coupling the theory can be
regarded as a gauged sigma-model. If the target $X$ is flat, the
theory constructed in \cite{GW} is superconformal and has
$SU(2)_N\times SU(2)_R$ R-symmetry with respect to which
supercharges transform as $(2,2)$. For general $X$, the model has
$SU(2)_N$ R-symmetry with respect to which the supercharges
transform as a doublet. In either case, one can twist $SU(2)_N$
R-symmetry to get a topological field theory. It is an unusual
theory, since it straddles the boundary between Schwarz-type and
Witten-type topological field theories. \footnote{In retrospect, it
is obvious that this dichotomy is not a good one: upon gauge-fixing
Chern-Simons theory \cite{Schwarz,WittenJones} becomes a Witten-type
topological theory. Rather, Schwarz-type theories are a subclass of
Witten-type theories distinguished by the property that they have a
unitary sector. In Chern-Simons theory this is the ghost-number zero
sector.}

In this paper we study the topologically-twisted version of the
Gaiotto-Witten theory. We find that in the case of flat $X$ it is
equivalent to the pure Chern-Simons theory
\cite{Schwarz,WittenJones} whose gauge group is a supergroup. More
precisely, the topologically-twisted Gaiotto-Witten theory is
obtained from the supergroup Chern-Simons theory by gauge-fixing the
odd part of the supergroup. The gauge group $G$ of the
Gaiotto-Witten theory is the residual (even) part of the supergroup.
This provides a simple explanation of the fact that Gaiotto-Witten
theories with flat $X$ are in one-to-one correspondence with Lie
superalgebras with nondegenerate invariant metric \cite{GW}. This
observation also helps to construct BRST-invariant Wilson loop
operators: they are naturally associated with finite-dimensional
representations of the supergroup.

For general $X$ the topologically twisted Gaiotto-Witten theory can
be regarded as a gauged version of the Rozansky-Witten theory (the
3d topological sigma-model with target $X$ constructed in
\cite{RozanskyWitten}). Thus it is a hybrid of Chern-Simons and
Rozansky-Witten theory. It is associated to a quadruple
$(G,\kappa,X,I)$, where $G$ is a compact Lie group, $\kappa$ is an
invariant metric on its Lie algebra, $X$ is a \HK\ manifold with a
tri-holomorphic action of $G$, and $I$ is a complex structure on $X$
such that the complex moment map with respect to the complex
symplectic form $\Omega_I$ is isotropic with respect to $\kappa$.

For a simple $G$, the metric on the Lie algebra is unique up to a
multiple. The most obvious choice of $X$ in this case is the
cotangent bundle of the flag variety $G_\CC/B$, where $B$ is a Borel
subgroup of $G_\CC$. It is well known that the complex moment map
for the obvious $G$-action on $T^*(G_\CC/B)$ is nilpotent. In fact,
the image of this map is precisely the set of nilpotent elements in
the Lie algebra of $G_\CC$. The moment map is generically one-to-one
and gives the so-called Springer resolution of the variety of
nilpotent elements.\footnote{We note that the Gaiotto-Witten model
with gauge group $G=SU(N)$ and target $T^*(G_\CC/B)$ previously
appeared in the string theory context \cite{GWSdual}. It is an $N=4$
$d=3$ field theory which describes the degrees of freedom living  on
the boundary of a stack of $N$ D3-branes ending on a a bound state
of $k$ NS fivebranes and one D5-brane. Here $k$ is the Chern-Simons
level, and it is assumed that the D3-brane theory is in the vacuum
where the gauge group $U(N)$ is broken down to its maximal torus
$U(1)^N$.}

Thus to any compact simple Lie group we can attach two natural 3d
TFTs with Chern-Simons terms: the ordinary Chern-Simons theory and
the Chern-Simons-Rozansky-Witten theory with target $T^*(G_\CC/B)$.
It is well known that the former theory is related to
representations of quantum groups. In this paper we begin the study
of the latter theory. Namely, we compute the algebra of local
operators (which turns out to be rather trivial, as in Chern-Simons
theory) and determine the category of Wilson loop operators. The
category of Wilson loops turns out to be a novel deformation of the
$G_\CC$-equivariant derived category of coherent sheaves on
$T^*(G_\CC/B)$. We show that such a deformation can be defined in a
rather general situation of a differential graded Poisson algebra
with a Hamiltonian action of a Lie group, provided the moment map
satisfies a certain constraint. In one special case we compute the
braiding of Wilson loops.

A. K. would like to thank Lev Rozansky and Sergey Arkhipov for
valuable discussions. This work was supported in part by the DOE
grant DE-FG03-92-ER40701.

\section{Construction of the Chern-Simons-Rozansky-Witten model.}

\subsection{Fields and BRST transformations}

In this section we will construct the CSRW theory ``from scratch'',
by postulating certain BRST transformations and then constructing a
suitable BRST-invariant Lagrangian. It is shown in the appendix that
the same theory can also be obtained by twisting the Gaiotto-Witten
theory.

Let $M$ be a Riemannian 3-manifold with local coordinates $x^\mu$,
$\mu=1,2,3.$ We are going to construct a gauged version of the
Rozansky-Witten model with target $X$, where $X$ is a \HK\ manifold
of complex dimension $\dim_{\CC}X=2n$ which admits an action of the
group $G$. Let $V_a$ for $a=1,\ldots,\dim\, G$ be the vector fields
on $X$ corresponding to this $G$-action. We can view them as
components of a section of $TX\otimes \frg^*$, where $\frg$ is the
Lie algebra of $G$. In local complex coordinates $\phi^{\hat
I}=(\phi^I,\phi^{\bar I})$ with $I,\bar I=1,\ldots,2n$ we write
these vector fields as
$$V_a=V_a^{\hat I}\partial_{\hat I}=V_a^I\partial_I +V_a^{\bar I}\partial_{\bar I}.$$
Thus under an infinitesimal $G$-transformation with parameters
$\eps^a$ the bosonic fields $\phi^I,\phi^{\bar I}$ transform as
$$
\delta_\eps\phi^I=\eps^a V_a^I,\quad \delta_\eps \phi^{\bar
I}=\eps^a V_a^{\bar I}.
$$
The vector fields $V_a$ satisfy
$$
[V_a,V_b]=f^c_{ab} V_c,
$$
where $f^c_{ab}$ are the structure constants of $\frg$.

In order for $G$ to be a global symmetry of the RW model it is
necessary and sufficient that for all $a$ the $(1,0)$ vector field
$V_a^I$ be holomorphic and preserve the symplectic structure
$\Omega_{IJ}$. We will further assume that the $G$ action preserves
the K\"ahler form on $X$. This implies that locally there exist
moment maps $\mu_+,\mu_-,\mu_3:X\raa\frg^*$, i.e. $\frg^*$-valued
functions on $X$ satisfying:
$$d\mu_{+~a}=-i_{V_a}(\Omega),\qquad d\mu_{-~a}=-i_{V_a}({\bar \Omega}),$$
$$d\mu_{3~a}=i_{V_a}(J).$$
Here  $\Omega=\half \Omega_{IJ}d\phi^I\wedge d\phi^J$ is the
holomorphic symplectic form, $J=ig_{I \bar K}d\phi^I\wedge
d\phi^{\bar K}$ the K\"ahler form on $X,$  and $i_{V}(\omega)$
stands for the inner product of a vector field $V$ with a form
$\omega.$ We will assume that the moment maps exist globally (this
is automatic if $X$ is simply-connected). The function $\mu_+$ is
holomorphic, while $\mu_-={\overline \mu_+}$ is antiholomorphic.
They satisfy
$$
\{\mu_{+a},\mu_{+b}\}=-f^c_{ab}\mu_{+c}.
$$
where the curly brackets are the Poisson brackets with respect to
the complex symplectic form $\Omega_{IJ}$. Similar formulas hold for
$\mu_-$ and $\mu_3$, with appropriate symplectic forms.

%i.e.$i_{V}J:=V^{\hat N}J_{{\hat N} {\hat M}}d\phi^{\hat M}.$

Another ingredient we need is a $G$-invariant nondegenerate
symmetric bilinear form on the Lie algebra $\frg$. We will denote it
$\kappa_{ab}$ and its inverse $\kappa^{ab}$. It satisfies
$$
\kappa_{ad} f^d_{bc}+\kappa_{bd} f^d_{ac}=0.
$$
Later we will subject $\kappa$ to an integrality constraint: its
restriction to the cocharacter lattice of $G$ (the lattice of
homomorphisms from $U(1)$ to the maximal torus of $G$) must be
integral.

Finally, we will require the moment map $\mu_+$ to satisfy
$$\mu_+\cdot\mu_+=0,$$
%identity \be \label{fundid} \quad \mu_+\cdot \mu_3=const,\quad
%2\mu_3\cdot \mu_3-\mu_+\cdot \mu_-=const\ee
where $\mu_{+}\cdot \mu_+=\kappa^{ab}\mu_{+a}\mu_{+b}$. We will see
that this is necessary for the BRST transformation to be nilpotent
on gauge-invariant observables.

The fields of the theory are
\begin{equation} \label{e2.1} \text{bosonic:}\quad\phi^{I},\phi^{\bar
I}, A_{\mu}^a\qquad\text{fermionic:}\quad \eta^{\bar
I},\chi_{\mu}^I.
\end{equation}
where $I,\, {\bar I}=1,\ldots,2n,\, \mu=1,2,3,\, a=1,\ldots \dim\,
G.$ $A^a_\mu dx^\mu$ is a connection 1-form on a principal
$G$-bundle $\sE$ over $M$. With respect to an infinitesimal gauge
transformation with a parameter $\eps^a(x)$ it transforms as
follows:
$$
\delta_\eps A^a=-\left(d\eps^a-f^a_{bc} A^b\eps^c\right)=-D\eps^a.
$$
Since the group $G$ acts on $X$, there is a fiber bundle over $M$
associated with $\sE$ and typical fiber $X$. Let us call it $X_\sE$.
The connection 1-form $A$ defines a nonlinear connection on $X_\sE$,
which locally can be thought of as a 1-form on $M$ with values in
the Lie algebra of vector fields on $X$. Concretely, this 1-form is
given by
$$
A^a V_a.
$$
Bosonic fields $\phi^I(x),\phi^{\bar I}(x)$ describe a section
$\phi$ of $X_\sE$. Their covariant differentials are defined by
$$
(D\phi)^I=d\phi^I+A^a V_a^I,\quad (D\phi)^{\bar I}=d\phi^{\bar
I}+A^a V_a^{\bar I}.
$$
The fermionic fields $\chi_\mu^I$ are components of a 1-form
$\chi^I$ on $M$ with values in $\phi^*(T_{X_\sE})$, where
$T_{X_\sE}$ is the $(1,0)$ part of the fiberwise-tangent bundle of
$X_\sE$. The fermionic field $\eta^{\bar I}$ is a 0-form on $M$ with
values in the complex-conjugate bundle $\phi^*({\overline
T}_{X_\sE})$.

BRST transformations of the fields are postulated to be
\begin{align}\label{BRSTQ}
\delta_{Q}\phi^{\bar I}&=\eta^{\bar I}, \\ \nn \delta_{Q}\phi^{I}&=0
\\ \nn
\delta_{Q}\eta^{\bar I}&=-{\bar \xi}^{\bar I},\\ \nn
\delta_{Q}\chi^{I}&=D\phi^{I}\\ \nn
\delta_Q
A^a&=\kappa^{ab}\chi^K\partial_K\mu_{+b}
\end{align}
where $D\phi^I=d\phi^I+A^a V_a^I$ and we define
$$\xi^I:=V^I\cdot \mu_-,\quad {\bar \xi}^{\bar I}:=V^{\bar I}\cdot \mu_+$$
Note that $\delta_Q^2$ is a gauge transformation with a parameter
$\eps^a=-\kappa^{ab}\mu_{+b}$:
$$\delta^2_Q A^a=\kappa^{ab}\left(d\mu_{+b}+f^d_{cb}A^c\mu_{+d}\right)$$
$$\delta^2_Q \phi^I=0,\quad \delta^2_Q \phi^{\bar I}=-V^{\bar
I}\cdot \mu_+$$
$$\delta^2_Q \chi^I=-\chi^J\p_J V^I\cdot \mu_{+},\quad
\delta^2_Q \eta^{\bar I}=-\eta^{\bar J}\p_{\bar J} V^{\bar
I}\cdot\mu_{+}.$$ To compute $Q^2$ we used
$V_a^K\Omega_{KJ}V_b^J=f^c_{ab}\mu_{+c}$ and $V^I\cdot \mu_+=0.$

The BRST differential is odd with respect to the $\ZZ_2$-grading
given by fermion number modulo $2$. In general, it is not possible
to promote this $\ZZ_2$-grading to a $\ZZ$-grading (i.e. to define a
$\ZZ$-valued ghost number so that $\delta_Q$ has ghost number $1$).
However, if $X$ has a $U(1)$ action which commutes with the
$G$-action and with respect to which $\Omega_{IJ}$ has charge $2$,
one can define a $U(1)$ ghost number symmetry as follows: its action
on $\phi^I,\phi^{\bar I}$ comes from the $U(1)$ action on $X$, while
the ghost numbers of fields $\chi,A,\eta$ are $-1,0,1,$
respectively. Taking into account that $\mu_+$ has ghost number $2$,
it is easy to check that $\delta_Q$ has ghost number $1$. This
situation occurs when $X$ is the cotangent bundle of a complex
manifold $Y$ and $U(1)$ acts multiplicatively on the fiber
coordinates.

\subsection{The classical action}
The BRST invariant action\footnote{We use Euclidean conventions for
the path-integral $Z=\int e^{-S}.$} consists of three parts:
$$ S={1\over \hbar}\int_M\, {\cal L},\quad {\cal L}={\cal L}_1+{\cal L}_2 +{\cal L}_{CS},
$$
where\footnote{The overall normalization of the Q-exact piece is
chosen for convenience.}
$${\cal L}_{CS}=\half \kappa_{ab}\Bigl( A^a \wedge dA^b-{1\over 3} f^b_{cd}A^a\wedge A^c \wedge A^d  \Bigr)
$$
$${\cal L}_1 = \delta_Q \Bigl(g_{I\bar K}\chi^I \wedge *D\phi^{\bar K}
-\sqrt{h}g_{I\bar K}\xi^I \eta^{\bar K}\Bigr)=g_{I\bar
K}\Bigl(D\phi^I \wedge *D\phi^{\bar K} -\chi^I\wedge \star
D\eta^{\bar K}\Bigr)+
$$
$$\sqrt{h} \Bigl(g_{I\bar K}\xi^I{\bar \xi}^{\bar K}+
g_{I \bar K}\partial_{\bar P}(\xi^I)\eta^{\bar K}\eta^{\bar
P}\Bigr)$$
$${\cal L}_2=\half \Omega_{IJ}\left(\chi^I\wedge
D\chi^J+\frac{1}{3}{\cal R}^J_{KL\bar
M}\chi^I\wedge\chi^K\wedge\chi^L\wedge\eta^{\bar M}\right).$$ Here
star denotes the Hodge star operator on forms on $M$ with respect to
a Riemannian metric $h_{\mu\nu}$, and covariant derivatives are
defined as:
$$D\phi^I=d\phi^I+A\cdot V^I,\quad D\chi^I=\nabla \chi^I +A\cdot \nabla_K(V^I)\chi^K,\quad
D\eta^{\bar I}=\nabla \eta^{\bar I} +A\cdot \nabla_{\bar K}(V^{\bar
I})\eta^{\bar K}$$ where $\nabla$ involves the Levi-Civita
connection on $X,$ i.e.
$$
\nabla\eta^{\bar J} =d\eta^{\bar J}+ \Gamma^{\bar J}_{\bar I \bar
K}d\phi^{\bar I}\wedge \eta^{\bar K},\quad \nabla\chi^{J}=d\chi^{J}+
\Gamma^{J}_{I K}d\phi^{I}\wedge\chi^{K}.
$$
(From now on, we will omit the sign $\wedge$ when writing the
exterior product of forms on $M$.) Finally, ${\cal R}^J_{KL\bar M}$
denotes the curvature tensor of the Levi-Civita connection on $X$:
$${\cal R}^{J}_{K L \bar M}={\p \Gamma^{J}_{K L}\over \p \phi^{\bar M}},\quad
\Gamma^I_{JK}=\left(\p_J g_{K \bar M}\right)g^{I \bar M}.$$

Gauge-invariance of the Chern-Simons action with respect to large
gauge transformations imposes a quantization condition on the
symmetric form $\kappa_{ab}/\hbar$. If $\kappa$ is chosen to be an
integral pairing on the cocharacter lattice of $G$, then the
quantization condition says
$$
\hbar=\frac{2\pi i}{k},\quad k\in\ZZ.
$$
The classical limit is $k\raa\infty$.

%Note that Q-trivial piece contains scalar potential
%$$V=J_{I\bar K}\xi^I {\bar \xi}^{\bar K}=f_{abc}\mu_+^a\mu_-^b\mu_3^c$$
%which is very useful in computing partition sum, see section 5
When checking the BRST-invariance of the action the following two
identities are useful: \be \label{impo}\left(\nabla_K \nabla_L
\partial_I \mu_+\right)\cdot \mu_+ + 3\partial_I\mu_+\cdot
\left(\nabla_K
\partial_L \mu_+\right)=0 \quad (IKL)\ee \be \label{impoii}
 \partial_{\bar M}\nabla_I \partial_K \mu_{3~a}=0\ee
where $(IKL)$ in (\ref{impo}) indicates symmetrization in indices
$I,K,L.$ The first one  follows from differentiating $\mu_+^2=0$ and
using that $\Omega_{IJ}$ is covariantly constant with respect to the
Levi-Civita connection. The second one follows from the definition
of $\mu_{3~a}$ and $\partial_I V_a^{\bar J}=0.$

The most non-trivial step in checking the BRST invariance of the
action is the cancelation of the two terms $O_1$ and $O_2$ arising
from $\delta_Q {\cal L}_2$:
$$O_1:=-\half \Omega_{IJ}\chi^I(\nabla_KV^J)\cdot (\delta_Q A)\chi^K$$
is canceled by
$$O_2:=-\frac{1}{6}\Omega_{IJ}{\cal R}^J_{KL\bar M}\chi^I \chi^K \chi^L (\delta_Q\eta^{\bar M})$$

To see this we first rewrite $O_1$ as
$$-\half \chi^I \chi^K \chi^L \left(\nabla_K \partial_I\mu_+\right)\cdot \partial_L\mu_+=
{1\over 6}\chi^I \chi^K \chi^L
\left(\nabla_K\nabla_L\partial_I\mu_+\right)\cdot \mu_+$$ where we
used (\ref{impo}). Then we further rewrite $O_1$ as
$${1\over 6}\chi^I \chi^K \chi^L \Omega_{IJ}\left(\nabla_K\nabla_L V^J\right)\cdot \mu_+=
-{1\over 6}\chi^I \chi^K \chi^L \Omega_{IJ}g^{J {\bar
M}}\left(\nabla_K\nabla_L\partial_{\bar M}\mu_3\right)\cdot \mu_+$$
Now we use
$$\nabla_K\nabla_L\partial_{\bar M}\mu_{3~a}=\partial_{\bar M}\nabla_K \partial_L \mu_{3~a}+
{\cal R}^P_{KL\bar M}\partial_P \mu_{3~a}$$ and (\ref{impoii}) to
bring $O_1$ to the form
$$O_1=-{1\over 6}\chi^I \chi^K \chi^L \Omega_{IJ}g^{J {\bar M}}{\cal R}^P_{KL\bar M}\partial_P \mu_{3}\cdot \mu_+=
-{1\over 6}\chi^I \chi^K \chi^L \Omega_{IJ}g^{J {\bar M}}{\cal
R}^P_{KL\bar M}g_{P{\bar N}}V^{\bar N}\cdot \mu_+$$ Lastly we
observe that for K\"ahler manifolds
$${\cal R}_{\bar N KL\bar M}:={\cal R}^P_{KL\bar M}g_{P{\bar N}}=\partial_{\bar M}\partial_{\bar N}\partial_K\partial_L {\cal K}
-\Gamma^P_{KL} \partial_{P}\partial_{\bar M}\partial_{\bar N}{\cal
K}$$ i.e. ${\cal R}_{{\bar N} KL {\bar M}}$ is symmetric in ${\bar
N},{\bar M}$ indices. In this way we obtain
$$ O_1=-{1\over 6}\chi^I \chi^K \chi^L \Omega_{IJ}{\cal R}^J_{KL\bar N}V^{\bar N}\cdot
\mu_+$$ and we conclude using $\delta_Q\eta^{\bar J}$ that
$$O_1+O_2=0.$$
\subsection{Gauge-fixing}
Next we discuss gauge-fixing in the CSRW model. This is somewhat
nontrivial, because even before gauge-fixing we have a BRST operator
$\delta_Q$. When we extend the theory by adding Faddeev-Popov
ghosts, anti-ghosts and Lagrange multiplier fields, we have to
define how $\delta_Q$ acts on them. The possibilities are actually
quite limited, since the total BRST operator
$$
\delta_\hQ=\delta_Q+\delta_{FP},
$$
must be nilpotent. Here $\delta_{FP}$ is the usual Faddeev-Popov
BRST operator. Note that the original BRST operator $\delta_Q$ is
nilpotent only up to a gauge transformation.

We introduce fermionic Faddeev-Popov ghost and anti-ghost fields
$c^a,{\bar c}_a,$ as well as bosonic Lagrange multiplier fields
$B_a$. We regard $c$ as taking values in $\frg$ and $\bar c$ and $B$
as taking values in $\frg^*$. The modified BRST operator ${\hat Q}$
acts as
$$\delta_{\hat Q}A_a=dc_a-f_{abd}A^bc^d+\chi^K\partial_K \mu_{+a}$$
$$\delta_{\hat Q}\phi^I=-V^{I}\cdot c,\quad
\delta_{\hat Q}\phi^{\bar I}=\eta^{\bar I}-V^{\bar I}\cdot c$$
$$\delta_{\hat Q}\chi^I=D\phi^I+(\partial_JV^{Ia})\chi^Jc_a$$
$$\delta_{\hat Q}\eta^{\bar I}=-{\bar \xi}^{\bar I}+
(\partial_{\bar J}V^{\bar Ia})\eta^{\bar J}c_a$$
$$\delta_{\hat Q}c^a=\kappa^{ab}\mu_{+b}-\half f^a_{bc} c^b c^c,\quad \delta_{\hat Q}{\bar c}=B,\quad
\delta_{\hat Q}B=0.$$ It is easy to check that $\delta_\hQ^2=0.$

One can express this result by saying that $B$ and $\bar c$ are
invariant under $\delta_Q$, while the ghost field is not:
$$
\delta_Q c^a=\kappa^{ab}\mu_{+ b}.
$$
The action of $\delta_{FP}$ on all fields is standard.

Let $f_a$ be a gauge-fixing function, then a $\delta_\hQ$-invariant
gauge-fixed action has the form
$$ S={1\over \hbar}\int_M\, {\cal L},\quad {\cal L}={\cal L}_1+{\cal L}_2 +{\cal L}_{CS},
$$
$${\cal L}_{CS}=\half \Bigl(\kappa_{ab} A^a \wedge dA^b-{1\over 3} f_{abc}A^a\wedge A^b \wedge A^c  \Bigr)
$$
$${\cal L}_1 = \delta_{\hat Q} \Bigl(g_{I\bar K}\chi^I \wedge *D\phi^{\bar K}
-\sqrt{h}g_{I\bar K}\xi^I \eta^{\bar K}+{\bar c}_af^a
\Bigr)=g_{I\bar K}\Bigl(D\phi^I \wedge *D\phi^{\bar K} -\chi^I\wedge
\star D\eta^{\bar K}\Bigr)+
$$
$$\sqrt{h} \Bigl(g_{I\bar K}\xi^I{\bar \xi}^{\bar K}+
g_{I \bar K}\partial_{\bar P}(\xi^I)\eta^{\bar K}\eta^{\bar
P}\Bigr)+B_a f^a -{\bar c}_a\delta_{\hat Q}f^a$$
$${\cal L}_2=\half \Omega_{IJ}\left(\chi^I\wedge
D\chi^J+\frac{1}{3}{\cal R}^J_{KL\bar
M}\chi^I\wedge\chi^K\wedge\chi^L\wedge\eta^{\bar M}\right).$$ Note
that the part of the action involving ghosts and anti-ghosts is not
standard, since it involves the $\delta_\hQ$-variation of the
gauge-fixing function rather than the usual $\delta_{FP}$ variation.
For example, if $f^a$ depends only on the gauge field (e.g. one
could pick the Lorenz gauge $f^a=\partial^\mu A^a_\mu$), the action
contains a term which couples ${\bar c}^a$ to the ``matter'' fermion
$\chi^K$.

\section{CSRW model for a flat target space}
\subsection{Relation to supergroup Chern-Simons theory}
It was shown in \cite{GW} that constraints of $N=4$ superconformal
symmetry amount to the following quadratic constraints on the moment
maps:
\begin{equation}\label{quadraticmoment}
\mu_{+}\cdot \mu_+=0,\quad \mu_3\cdot \mu_+=0,\quad
2\mu_3\cdot\mu_3-\mu_+\cdot\mu_-=0.
\end{equation}
Equivalently, if we define a 3-vector of moment maps $\mu_{i\,
a}=(\mu_{1a},\mu_{2a},\mu_{3a})$ by letting
$$
\mu_+=\mu_1+i\mu_2,\quad \mu_-=\mu_1-i\mu_2,
$$
then the constraint says that the traceless part of the symmetric
tensor
$$
K_{ij}=\kappa^{ab}\mu_{i\,a}\mu_{i\,b}
$$
vanishes.

In the case when $X$ is a vector space with a linear action of $G$
the functions $\mu_{ia}$ are quadratic. For example, $\mu_{+a}=\half
\kappa_{ab}\tau^b_{IJ}\phi^I\phi^J$, where $\tau^b_{IJ}$ are
constants. It was noted in \cite{GW} that the quadratic constraints
on $\mu_{ia}$ are equivalent to the requirement that $\tau_{a IJ}$
together with $f^a_{bc}$ are structure constants of a Lie
superalgebra whose even part is $\frg$ and odd part is $X$.

This connection of $N=4$ $d=3$ superconformal field theories with
Lie superalgebras seems quite mysterious. In this section we
demystify it to some extent. We show that for flat $X$ the
topologically twisted Gaiotto-Witten theory is the supergroup
Chern-Simons theory with a partial gauge-fixing (the odd part of the
supergroup gauge-invariance is fixed, while the even one is not).
The BRST differential $\delta_Q$ arises from such a partial
gauge-fixing. Upon gauge-fixing the residual bosonic gauge symmetry,
the twisted Gaiotto-Witten model becomes the supergroup Chern-Simons
theory with the usual Faddeev-Popov gauge-fixing. The bosonic fields
$\phi^I,\phi^{\bar I}$ are interpreted as bosonic Faddeev-Popov
ghosts and anti-ghosts corresponding to odd gauge symmetries.

We begin by recalling that for flat $X$ the moment maps take the
form \be \label{flatmom}\mu_{+a}=\half\kappa_{ab} \tau^b_{IJ}\phi^I
\phi^J,\quad \mu_{- a}= \half \kappa_{ab}\tau^b_{\bar I \bar
J}\phi^{\bar I} \phi^{\bar J},\quad \mu_{3 a}=-{i}\kappa_{ab}
\phi^I\tau^b_{IJ}\Omega^{JK}g_{K \bar M}\phi^{\bar M},\ee where
$\tau^a_{\bar I \bar L}=\Omega_{\bar I \bar P }g^{\bar P M}\tau^a_{MJ}
\Omega^{JK}g_{K \bar L}$ and
$\Omega^{IJ}=-{1\over 4}g^{I \bar K}\Omega_{\bar K \bar M}g^{\bar M
J}$ is the inverse of $\Omega_{IJ}.$

Following \cite{GW}, we introduce a Lie superalgebra $\frG$ whose
even part is $\frg$ and odd part is the vector space $X$. Let $M_a$
be a basis in $\frg$ and $\lambda_J$ be a basis in $X$ dual to
coordinate functions $\phi^J$. Then the commutation relations of
$\frG$ are defined to be
$$
[M_a,M_b]=f_{ab}^cM_c,\quad
[M_a,\lambda_I]=\kappa_{ab}\tau^b_{IJ}\Omega^{JK}\lambda_K,\quad
\{\lambda_I,\lambda_J\}=\tau^a_{IJ}M_a.
$$
The super-Jacobi identities are equivalent to
(\ref{quadraticmoment}). Note also that $\frG$ has a natural
super-trace (i.e. nondegenerate $ad$-invariant graded-symmetric
bilinear form):
$$\STr(M_aM_b)=\kappa_{ab},\quad \STr(\lambda_I
\lambda_J)=\Omega_{IJ}.$$ This enables one construct a Chern-Simons
action based on a supergroup associated to $\frG$. We will call it
super-Chern-Simons theory.

It turns out that for flat $X$ one can rewrite the action of the
CSRW model as the action of the super-Chern-Simons theory with
gauge-fixed odd part of the gauge symmetry. To see this, we
introduce the following fields with values in $\frG$: \be
\label{csfields} {\cal A}={\cal A}_b+{\cal A}_f,\quad {\cal
A}_b=A^aM_a,\quad {\cal A}_f=\chi^I \lambda_I,\ee
\be\label{csfieldsii}{\overline C}=\phi^{\bar I}g_{\bar I
K}\Omega^{KJ}\lambda_J,\quad C=\phi^I\lambda_I,\quad B=\eta^{\bar
M}g_{{\bar M} K}\Omega^{KI}\lambda_I. \ee Here $C$ and $\bar C$ are
Faddeev-Popov ghosts and anti-ghosts for the fermionic gauge
symmetry, while $B$ is a fermionic Lagrange multiplier field.

The BRST operator $\delta_Q$ of the CSRW model can be interpreted as
arising from gauge-fixing the fermionic part of the gauge symmetry.
in super-Chern-Simons theory. In terms of the fields
(\ref{csfields}),(\ref{csfieldsii}) $\delta_Q$ acts as follows:
$$\delta_Q {\cal A}=dC-[{\cal A},C\},\quad \delta_Q {\overline
C}=B,\quad \delta_Q C=0,$$
$$\delta_Q B=\half [{\overline C},[C,C\}\}.$$
The symbol $[~\}$ stands for the graded commutator in the Lie
algebra.

The first three of these transformation laws are standard. (The BRST
variation of $C$ vanishes because we introduced ghosts only in the
odd part of $\frG$). The transformation law of the Lagrange
multiplier field $B$ is unusual: in the Faddeev-Popov gauge-fixing
procedure the BRST-variation of the Lagrange multiplier field is
zero. This difference arises because the odd part of $\frG$ is not a
Lie subalgebra. For this reason, $\delta_Q$ is not nilpotent when
acting on $A$, but satisfies
$$
\delta_Q^2 A^a=\kappa^{ab}D\mu_{+ b}.
$$
This is a gauge transformation with respect to the residual gauge
symmetry with a parameter $-\kappa^{ab}\mu_{+b}$. Consistency
requires that $\delta_Q^2$ act as a gauge transformation on all
fields, and this determines the BRST variation of $B$.

We note in passing that one can do partial gauge-fixing in ordinary
Yang-Mills theory with purely bosonic gauge symmetry. For example,
one can consider gauge-fixing the off-diagonal part of $SU(2)$ gauge
symmetry in an $SU(2)$ Yang-Mills theory. The corresponding BRST
operator will not be nilpotent because the part of the gauge
symmetry that we fix is not a subalgebra of the full gauge symmetry.
Rather, its square will be a gauge transformation with respect to
residual $U(1)$ gauge symmetry. Of course, once we gauge-fix the
residual $U(1)$ symmetry, we recover the usual theory with a
nilpotent BRST operator.

In terms of $\frG$-valued fields the action of the CSRW model takes
the form \be \label{flatact} S=\hbar^{-1}\int_M {\cal L},\quad {\cal
L}=\half \STr \Bigl({\cal A}\,d{\cal A} - {1\over 3}{\cal A}[{\cal
A},{\cal A}\}\Bigr)-\delta_Q \Psi\ee where the gauge-fixing fermion
$\Psi$ is taken to be
$$\Psi=
 \STr\Bigl({\cal
A}_f\star \bigl(d{\overline C}-[{\cal A}_b, {\overline C}\}\bigr)-2
\vol_M \STr\Bigl([\bar C,\bar C\}[C,B\}\Bigr).$$ This shows that for
flat $X$ the CSRW model is equivalent to super-Chern-Simons theory
with the fermionic part of the gauge symmetry fixed.

As discussed in Section 2.3, we can gauge-fix the remaining bosonic
gauge symmetry in the CSRW model by introducing the fermionic fields
$c^a,{\bar c}_a$ and bosonic fields $B_a$ and modifying the BRST
operator $\delta_Q$. In the case of flat $X$ it is illuminating to
introduce the fields taking values in $\frG$:
$${\bf C}=c^aM_a+\phi^I\lambda_I,\quad {\bar {\bf C}}={\bar c}^aM_a+
\phi^{\bar I}g_{\bar I K}\Omega^{KJ}\lambda_J$$
$${\bf B}=B^aM_a+(\eta^{\bar I}-V^{\bar I}\cdot c)g_{\bar I
K}\Omega^{KJ}\lambda_J.$$ In terms of these fields the modified BRST
differential $\delta_\hQ$ acts as follows:
$$\delta_{\hat Q}{\cal A}=d{\bf C}-[{\cal
A},{\bf C}\},\quad \delta_{\hat Q} {\overline {\bf C}}={\bf B},\quad
\delta_{\hat Q} {\bf C}=-\half [{\bf C},{\bf C}\},\quad \delta_{\hat
Q} {\bf B}=0.$$ These are the usual BRST transformations laws. The
nilpotency of the operator $\delta_\hQ$ follows immediately from the
Jacobi identities for $\frG$. Note that the new Lagrange multiplier
field ${\bf B}$ is BRST-invariant, as it should. In terms of the new
fields the gauge-fixed action of the CSRW model is identical to the
fully gauged-fixed super-Chern-Simons action for a particular choice
of the gauge-fixing fermion: \be \label{flatactiii}
S_{g.f.}=\hbar^{-1}\int_M {\cal L},\quad {\cal L}=\half \STr
\Bigl({\cal A}\,d{\cal A} - {1\over 3}{\cal A}[{\cal A},{\cal
A}\}\Bigr)-\delta_\hQ {\hat\Psi}\ee where
\begin{equation}\label{hatPsi}
{\hat\Psi} = \STr\Bigl({\cal A}\star \bigl(d{\overline {\bf
C}}-[{\cal A}, {\overline {\bf C}}\}\bigr)\Bigr) -2 \vol_M
\STr\Biggl([\bar {\bf C},\bar {\bf C}\}[{\bf C},\Bigl({\bf
B}+[{\overline {\bf C}},{\bf C}\}\Bigr)\}\Biggr)
\end{equation}
The only unusual thing about it is the form of the gauge-fixing
fermion (usually it is taken to be independent of ghost and
anti-ghost fields). These ghost-dependent terms are introduced to
reproduce the scalar potential $V=\xi^I g_{I \bar J}{\bar \xi}^{\bar
J}$ and the fermionic mass term $g_{I\bar K}\p_{\bar
J}(\xi^I)\eta^{\bar K}\eta^{\bar J}$ present in the CSRW model.

\subsection{Local observables}
To find  topological observables, we have to compute the cohomology
of the BRST operator ${\hat Q}$ in the space of 0-forms. In
Chern-Simons theory with a compact gauge group one usually restricts
to observables with ghost number $0$. Then the only observable is
the identity operator, basically because there is nothing to
construct the local observable from other than the Faddeev-Popov
ghosts. If we impose the same restriction in the super-Chern-Simons
theory, we get the same trivial result, for exactly the same reason.

However, if we regard the CSRW model as a gauged version of the
Rozansky-Witten model, it seems unreasonable to restrict oneself to
observables of ghost number $0$. In ordinary Chern-Simons theory the
ghost-number zero sector is distinguished by its unitarity
properties, but this is no longer the case in the CSRW model.

Candidate observables in the super-Chern-Simons theory are
polynomial functions of the $\frG$-valued field ${\bf C}$. Its BRST
transformation is
$$
\delta_\hQ {\bf C}=-\frac12[{\bf C},{\bf C}\}.
$$
This is simply the Chevalley-Eilenberg differential in the complex
which computes the cohomology of $\frG$ with trivial
coefficients.\footnote{Similarly, if we allowed ghost-dependent
observables in the ordinary Chern-Simons theory with gauge group
$G$, we would get the cohomology of $\frg$ with trivial coefficients
as the answer.} For $\frG=\gl(m\vert n)$ and $\frG=\osp(m\vert n)$
this cohomology is finite-dimensional \cite{Fuks}. For example, for
$\gl(m\vert n)$ it is isomorphic to the cohomology of the bosonic
Lie algebra $\gl({\rm max}(m,n))$ \cite{Fuks}. In particular, the
cohomology of $\gl(1\vert 1)$ is isomorphic to the exterior algebra
with one generator of ghost number one.

From the point of view of the Gaiotto-Witten theory, this result may
seem somewhat surprising, since an arbitrary $G$-invariant
holomorphic function of $\phi^I$ is obviously $\delta_\hQ$-closed.
However, because of nonstandard transformation properties of the
$c^a$ ghosts, all such observables are $\delta_\hQ$-exact. For
example, for $\frg=\gl(1\vert 1)$ ($G=U(1)\times U(1)$) we have two
complex scalars $A,B$ with $U(1)\times U(1)$ charges $(1,-1)$ and
$(-1,1)$, so the only gauge-invariant holomorphic function is $AB$.
But this is proportional to $\delta_\hQ (c_1+c_2)$, where $c_1$ and
$c_2$ are the Faddeev-Popov ghosts for the two $U(1)$ factors.

Apart from ordinary local observables built as polynomials in the
fields, there may also be local observables which are disorder
operators. These are monopole operators, i.e. operators which insert
a Dirac monopole singularity at a point \cite{BKW1,BKW2}. Here we
limit ourselves to the simplest case $G=U(1)\times U(1)$,
$\frG=\gl(1\vert 1)$. This case is simple because the bosonic part
of $\frG$ is abelian. The monopole sits in the bosonic part of the
gauge group and is characterized by the property that the gauge
field strengths $F_1$ and $F_2$ are singular at the insertion point:
$$
F_1=* d\frac{m_1}{2r}+{\rm regular},\quad F_2=* d\frac{m_2}{2r}+{\rm
regular},
$$
where $r$ is the distance to the insertion point and $m_1,m_2$ are
integers (magnetic charges). We will denote the corresponding
operator $M_{m_1,m_2}$. In the presence of such a singularity the
Chern-Simons action is not gauge-invariant:
$$
\delta_\eps S_{CS}=ik \left(m_1 \eps_1-m_2\eps_2\right),
$$
where $\eps_1,\eps_2$ are the parameters of the $U(1)\times U(1)$
gauge transformation. Thus the monopole operator has electric
charges $km_1, -km_2$. Upon gauge-fixing, this translates into the
following transformation law of $M_{m_1,m_2}$ under the
BRST-transformation:
$$
\delta_\hQ M_{m_1,m_2}=-ik \left(m_1 c_1-m_2 c_2\right) M_{m_1,m_2}.
$$

To get a BRST-invariant operator, one must multiply $M_{m_1,m_2}$ by
a suitable polynomial in the ghost fields (both bosonic and
fermionic). Since $\delta_\hQ$ variations of all these fields do not
involve $c_1+c_2$, it is obvious that a necessary condition for the
existence of a suitable polynomial is $m_1=m_2=m$. Then for $km>0$
the BRST-invariant combination has the form
$$
 f(A,B,c_1,c_2)B^{km}M_{m_1,m_2},
$$
while for $km<0$ it has the form
$$
 f(A,B,c_1,c_2)A^{-km} M_{m_1,m_2}.
$$
Here $f(A,B,c_1,c_2)$ is an arbitrary BRST-invariant function of
$A,B,c_1,c_2$. We also must identify functions which differ by
BRST-exact terms. This is exactly the same problem as what we
encountered above when computing the space of ``ordinary'' local
observables. As discussed above, this space is isomorphic to the
cohomology of the bosonic Lie algebra $\gl(1)$, which is an exterior
algebra with one generator with ghost number one. This generator is
$c_1+c_2$. Thus for every nonzero magnetic charge the space of
BRST-invariant monopole operators is two-dimensional and has a
one-dimensional even subspace and one-dimensional odd subspace. The
even component has ghost number $|km|$, while the odd component has
ghost number $|km|+1$.

For $m=0$ the situation is similar: $\delta_\hQ$-cohomology is
spanned by $1$ and $c_1+c_2$. As mentioned above, this is
interpreted as the cohomology of $\frG$ with trivial coefficients.

To summarize, the space of local observables in the CSRW model with
gauge group $U(1)\times U(1)$ at level $k$ is the tensor product of
the cohomology of $\frG$ (which is isomorphic to the exterior
algebra with one generator of ghost number $1$) and an
infinite-dimensional vector space $V$ graded by the magnetic charge
$m\in\ZZ$. We will refer to the two factors in this tensor product
as perturbative and nonperturbative spaces. For any $m$ the
component $V_m$ of the nonperturbative space is one-dimensional and
has ghost number $|km|$.

The space of local observables in a 3d TFT must be an associative
graded-commutative algebra. For $G=U(1)\times U(1)$ it is easy to
determine the algebra structure. First of all, it is clear that the
factorization into perturbative and nonperturbative parts persists
on the algebra level. The perturbative algebra is the exterior
algebra with one generator. The nonperturbative algebra is tightly
constrained by the conservation of magnetic charge: the product of
two monopole operators with magnetic charges $m,n$ is a monopole
operator with magnetic charge $m+n$. Further, since $AB$ is
$\delta_\hQ$-exact, the product of monopole operators with $mn<0$ is
$\delta_\hQ$-exact. Thus the nonperturbative algebra is a
commutative algebra with three generators $1,x,y$ and relations
$$
1\cdot x=x,\quad 1\cdot y=y,\quad x\cdot y=0.
$$
Both generators $x,y$ have ghost number $|k|$, while their magnetic
charges are $\pm 1$.

\section{CSRW model for a curved target space}
In this section we consider the CSRW model for a curved target
space. We consider in detail the case $G=SU(2)$, $X=T^*\CC\PP^1$.
Then we discuss a generalization where $G$ is an arbitrary compact
simple Lie group and $X$ is the cotangent bundle of the flag
manifold $G_{\CC}/B$.

\subsection{Moment maps}

The cotangent bundle of any complex manifold is a complex symplectic
manifold. If $x^i$ are local coordinates on the base and $p_i$ are
dual coordinate on the fibers, then the symplectic form is simply
$$
\Omega=dp_i dx^i.
$$
If the base manifold admits a holomorphic $G_\CC$ action, then
$G_\CC$ acts on the total space of the cotangent bundle and this
action is Hamiltonian. The corresponding complex moment maps are
simply
$$
\mu_{+a}=v^i_a p_i
$$
where $v_a=v^i_a \partial_i$ is the holomorphic vector field on the
base corresponding to a basis vector $e_a\in\frg_\CC$. It is more
difficult to satisfy the condition $\kappa^{ab}\mu_{+a}\mu_{+b}=0$:
this requires the vector fields $v_a$ to be null with respect to the
metric $\kappa$.

One simple situation where this happens is when the base manifold is
itself the quotient of $G_\CC$ by a Borel subgroup, i.e. the flag
manifold of $G$. The simplest nontrivial case is $G=SU(2)$, in which
case the flag manifold is $\CC\PP^1$. It is well-known that
$T^*\CC\PP^1$ admits a \HK\ metric known as the Eguchi-Hanson metric
(see e.g. \cite{EGH}). The group $SU(2)$ acts on it by isometries
and preserves all three complex structures. Below we summarize some
properties of this manifold and of the $SU(2)$ action on it.

Let $z$ be an inhomogeneous coordinate on $\CC\PP^1$ and $b$ be a
complex coordinate on the fiber of $T^*\CC\PP^1$. The
complexification of $SU(2)$ is $SL(2,\CC)$ and it acts as follows:
\be \label{transf} z\mapsto {\alpha z+\beta\over \gamma
z+\delta},\quad b\mapsto b (\gamma z +\delta)^2\ee Let us introduce
$(0,1)$ forms
$$e_1=\half {db \over b}+{\bar z}e_2,\quad e_2={dz\over 1+\vert z\vert^2}$$
such that $e_1$ and $e_2 \sqrt{b\over {\overline b}}$ are invariant
under $SU(2).$ The K\"ahler form on $T^*\CC\PP^1$ which respects the
$SU(2)$ symmetry is \be J=t{\hat J},\quad {\hat
J}=i\Bigl(f_1(x)e_1\wedge {\bar e}_1+f_2(x)e_2\wedge {\bar
e}_2\Bigr). \label{metric} \ee Here $t$ sets the scale, $\int_{{\bf
P}^1}{\hat J}=2\pi$ and \be f_1(x)={x^2\over f_2},\quad
f_2(x)=\sqrt{1+x^2} \label{metricii} \ee are functions of the only
$SU(2)$ invariant
$$x^2=\vert b\vert^2(1+\vert z\vert^2)^2.$$
A harmonic holomorphic $SU(2)$-invariant (0,2) form $\Omega$ which
respects $SU(2)$ symmetry is given by \be \label{holo}\Omega=t{\hat
\Omega},\quad \hat \Omega=db \wedge dz\ee Note that theory does not
dependent on $t$ since all $t$-dependence in the BRST-non-exact
piece of the action can be absorbed into rescaling $\chi$ and
$\eta.$
We note as an aside that the relation between our coordinates $z$
and $b$ and the standard coordinates $r,\theta,\phi,\psi$ on the
Eguchi-Hanson space \cite{EGH} is
$$x=r^2(1-r^{-4})^{1\over 2},\quad z={sin \theta\over 1-cos{\theta}}e^{i\phi},\quad
b={xe^{i(\phi-\psi)}\over 1+\vert z\vert^2}$$

From the transformation (\ref{transf}) we find the vector fields \be
\label{vectors}
V^z=(-i)\left(\begin{tabular}{cc}$-z$ & $1$\\ $-z^2$ & $z$\\
\end{tabular}\right),\quad
V^{b}=(-i)b\left(\begin{tabular}{cc}$1$ & $0$\\ $2z$ & $-1$\\
\end{tabular}\right)
\ee
$$
V^{\bar z}=i\left(\begin{tabular}{cc}$-{\bar z}$ & $-{\bar z}^2$\\ $1$ & ${\bar z}$\\
\end{tabular}\right),\quad
V^{\bar b}=i {\bar b}\left(\begin{tabular}{cc}$1$ & $2{\bar z}$\\ $0$ & $-1$\\
\end{tabular}\right)
$$
Using (\ref{metric}) and (\ref{holo}) we find the moment maps
$\mu=t{\hat \mu}$ with \be \label{moments} {\hat \mu}_+=(-i)b\left(
\begin{tabular}{cc}
$-z$ & $ 1 $\\
$-z^2$ & $z$\\
\end{tabular}
\right),\quad {\hat \mu}_-=i{\bar b}\left(
\begin{tabular}{cc}
$-{\bar z}$ & $ -{\bar z}^2$\\
$1$ & ${\bar z}$\\
\end{tabular}
\right) \ee

\be {\hat \mu}_3={f_2(x) \over 1+\vert z\vert^2}\left(\begin{tabular}{cc}$\half(1-\vert z\vert^2)$ & ${\bar z}$\\
$z$ & $-\half(1-\vert z\vert^2)$\\
\end{tabular}\right)
\label{realmom} \ee

For $G=SU(2)$ the metric on the Lie algebra is given in terms of the
trace $\Tr$ so that moment maps satisfy:
\begin{equation}\label{momentstrace}
\Tr\Bigl(\mu_+^2\Bigr)=0,\quad \Tr\Bigl(\mu_+
\mu_3\Bigr)=0,\quad 2\Tr\Bigl(\mu_3^2\Bigr)-\Tr\Bigl(\mu_+
\mu_-\Bigr)=t^2.
\end{equation}

\subsection{Local observables}

In this section we compute the algebra of local observables. We will
see that it is two-dimensional and isomorphic to the cohomology of
the Lie algebra $\sl(2)$.

We begin with local observables which do not depend on the fermionic
ghosts $c^a$. Such observables can be regarded as $(0,p)$ forms on
the target space. Such a form $\omega$ is annihilated by
$\delta_\hQ$ if an only if it is $SU(2)$-invariant and is
annihilated by
$$Q={\overline \partial}+i_{{\bar \xi}}$$
and the  where
$${\bar \xi}={x^2\over {\overline b}}\partial_{\bar z}-{2x^2 z \over 1+\vert z\vert^2}\partial_{\bar b}.$$

Suppose first that $\omega$ is odd, i.e. it is a $(0,1)$ form. The
requirement of $SU(2)$-invariance restricts its form to be
$$\omega_{odd}=C_1(x){\bar e}_1+C_2(x){\bar e}_2 \sqrt{{\overline b}\over b}.$$
We find that $Q\omega_{odd}=0$ implies $C_2(x)=0$ and does not
restrict $C_1(x).$ Note that $C_1(x)$ should  be an even function of
$x$ and vanish at least as $x^2$ for $x\mapsto 0$  so that
$\omega_{odd}$ is smooth. However, all such $\omega_{odd}$ are
$Q$-exact since
$$C_1(x){\bar e}_1={\overline \partial} F(x),\quad C_1(x)=xF'(x)$$
with smooth $F(x),$ vanishing at least as $x^2$ for $x\mapsto 0.$

Consider now an even form
$$\omega_{even}=A(x)+B(x)d{\bar b} \wedge d{\bar z}.$$
Here $A(x)$ and $B(x)$ are even functions of $x$ which are smooth at
$x\mapsto 0$ so that $\omega_{even}$ is also smooth. We find that
$Q\omega_{even}=0$ implies $A'(x)=2xB(x).$ However,
$$\omega_{even}=Q(\Upsilon),\quad \Upsilon=c(x){\bar e}_2 \sqrt{{\overline b}\over b}$$
where
$$A(x)=xc(x),\quad B(x)={1\over 2}(c'(x)+{c(x)\over x})$$
$\Upsilon$ is well behaved if and only if $c(x)$ is an odd function
of $x$ and $c(x)\mapsto x^{1+2m},\, m\ge 0$ for $x\mapsto 0.$
Therefore only the solution with $A(x)=1$ is not BRST-trivial. We
conclude that the BRST cohomology is one-dimensional corresponding
to $A(x)=1,\, B(x)=0.$

If the fermionic ghosts transformed in the standard way, we could
conclude from here that apart from $1$ the only BRST cohomology
classes are the ones constructed from the fermionic ghosts $c^a$
alone. That is, the cohomology is isomorphic to the cohomology of
the Lie algebra $\sl(2)$. This cohomology has a single generator in
degree three
$$
\eps_{abc} c^a c^b c^c.
$$
In our case, the BRST transformation of $c^a$ is nonstandard, but
this does not affect the structure of the cohomology. Indeed, we can
view the difference between $\delta_\hQ(c^a)$ and the ordinary
Chevalley-Eilenberg differential acting on $c^a$ as a perturbation
and write down a spectral sequence which converges to the desired
answer and whose $E_1$ term is the cohomology of $\sl(2)$. But since
the only nonzero terms in the $E_1$ term are of ghost number $0$ and
$3$, there can be no nontrivial differentials and the spectral
sequences collapses at the very first stage.\footnote{The concrete
form of the nontrivial class of degree $3$ is of course affected by
the perturbation: it is easy to check that the $\hQ$-closed
combination is $$\frac{1}{3}\eps_{abc} c^a c^b c^c-\mu_{+a} c^a$$.}

Unlike the case of flat $X$, there are no BRST-invariant monopole
operators in this model. To see this, let us adopt the radial
quantization viewpoint, i.e. let us consider the space of states of
theory on a manifold of the form $S^2\times\RR$ where $\RR$ is
regarded as time. Monopole singularity corresponds to a constant
magnetic flux on $S^2$. This flux breaks the gauge group down to
$U(1)$. Thanks to the Chern-Simons term the monopole operator has
electric charge with respect to this unbroken $U(1)$, and this
charge must be canceled by zero modes of other fields. However,
there are no such zero modes. The bosonic field corresponding to the
target coordinate $b$ is massive, while the field $z$ gets an
effective potential from its interaction with a constant backround
magnetic flux (the zeroes of this potential are the two points on
$\CC\PP^1$ which are fixed by the unbroken $U(1)$). The fermionic
fields $\eta^{\bar I}$ are also massive. Hence no BRST-invariant
monopole operators are possible.

\subsection{Generalization to an arbitrary gauge group}
The discussion can be easily generalized to any compact simple Lie
group $G$ and $X=T^*(G_\CC/B)=T^*(G/T)$ where $T$ is a maximal torus
of $G.$ Let us denote $r=\dim~T$ and expand the gauge field in the
Cartan-Weyl basis of $\frg$:
$$A=A^iH_i+\sum_{\alpha \in {\cal S}} \bigl(A^{\a}E_{\a}+A^{\bar \a}E_{\bar \a}\bigr)$$
where $\cal S$ is the set of positive roots. Let $w^{\a}$ be  complex
coordinates on the coset $G/T$ near the identity element of $G$:
$$\rho=e^{i\sum_{\a \in \cal S}\bigl(w^{\a}E_{\a}+{\bar w}^{\bar \a}E_{\bar \a}\bigr)}h,\quad
h\in T$$ It is convenient to work with holomorphic Darboux
coordinates on $X,$ $\phi^I=(b_{\a},z^{\alpha})$ for $\a \in \cal
S,$ so that the holomorphic symplectic form is
$$\Omega=t\sum_{\a \in \cal S}db_{\a}\wdg dz^{\a}.$$
Near the unit of $G,$ for $\vert w_{\a}\vert \ll 1,$ we find
$z^{\a}=w^{\a}+O(\vert w\vert^3).$ Similar to the previous
discussion for $T^*\CC\PP^1$ we introduced an overall scale factor
$t$ which enters both the holomorphic symplectic form $\Omega=t{\hat
\Omega}$ and the K\"ahler form $J=t{\hat J}.$

The group $G$ acts on the right coset from the left:
$$g:\rho\mapsto g \rho ,\quad g\in G.$$
As explained above, the holomorphic moment map is linear in the
fiber coordinates $b_{\a}$: \be \label{muplusgen} \mu_+=t {\hat
\mu}_+,\quad \hat \mu_+=\sum_{\a \in \cal S}b_{\a}V^{z^{\a}}(z)\ee
where \be \label{vectorgen} V^{z^{\a}}=E_{\a}+iz^{\a}\sum_{j=1}^r
\a^j H_j +\sum'_{\delta}z^{\a -\delta} E_{\delta}+O(z^2)\ee
%z^{\a}\sum_{\delta}z^{\delta}(\a \cdot \delta)E_{\bar \delta} +i\sum_{j=1}^r
%\sum'_{\delta}\delta^jH_jz^{\delta}z^{\a-\delta}\ee
Note that $G/T$ is a homogenous space so it is sufficient to work
near the identity element of $G,$ i.e. for $\vert z\vert \ll 1.$ In
(\ref{vectorgen}) $\sum'_{\delta}$ means that we sum over $\delta
\in \cal S$ such that $\a -\delta \in \cal S.$

From (\ref{muplusgen}) we find
$$V^{b_{\a}}=\sum_{\b \in \cal S} \partial_{z^{\a}}V^{z^{\b}},\quad
\xi^{z^{\a}}=t\sum_{\gamma} N^{\a {\bar \gamma}}{\bar b}_{\bar
\gamma},\quad \xi^{b_{\a}}=t\sum_{\gamma}\sum_{\b} {\bar b}_{\bar
\gamma}b_{\b}\partial_{z^{\a}}N^{\b {\bar \gamma}}$$ where the
matrix
$$N^{\a \bar \gamma}=\Tr\Bigl(V^{{\bar z}^{\bar \gamma}} V^{z_{\a}}\Bigr)$$
is non-degenerate. The nondegeneracy can be most easily seen at
$z=0$, and then must also be true in an open neighborhood of $z=0$.

An important consequence of the nondegeneracy of the matrix $N$ is
that the fields $b_\a$ and the fermions $\eta^{\bar I}$ are massive.
Thus the only zero modes are the bosons parameterizing the base
$G_\CC/B$. This implies that when the theory is considered on a
manifold $S^2\times\RR$ (without magnetic flux) the BRST cohomology
can be computed in the space of holomorphic functions on $G_\CC/B$
tensored with the zero modes of the Faddeev-Popov ghosts. Since
$G_\CC/B$ is compact, the only holomorphic function is a constant,
and we conclude that the BRST cohomology is isomorphic to the
cohomology of the Lie algebra $\frg$ with trivial coefficients.
Monopole operators do not arise for the same reason as for
$G=SU(2)$.

\section{Wilson loops in the CSRW model}
\subsection{General construction}
In Chern-Simons theory the most important observables are the Wilson
loops. Their correlators are known to give knot invariants
associated with representation theory of quantum groups. Similarly,
we expect correlators of Wilson loops in the CSRW model to compute
some knot invariants. While it is very plausible that these
invariants are also related in some way to quantum groups, the
precise relationship is unclear at the moment.

In the Chern-Simons theory the Wilson loops are labeled by
finite-dimensional irreducible representations of $G$.\footnote{This
statement is true in the classical limit. For finite $k$ it one
wants to preserve unitarity one has to keep only the so-called
integrable representations, i.e. those representations which can be
deformed into irreducible representations of the quantum group.} In
the the Rozansky-Witten model they are labeled by objects of the
$\ZZ_2$-graded derived category of coherent sheaves on $X$. A
natural guess is that Wilson loops for the CSRW model are labeled by
objects of the $G_\CC$-equivariant derived category of $X$. We will
see in this section that this is almost correct: the relevant
category is a certain interesting deformation of the
$G_\CC$-equivariant derived category of $X$. The existence of this
deformation is tied to the fact that the $G_\CC$-action is
Hamiltonian and the moment map is isotropic.

Let $E=E^++E^-$ be a $\ZZ_2-$graded vector bundle over $X$. Its
fiber over a point $p\in X$ is interpreted as the space of degrees
of freedom living on the Wilson loop. Observables on the Wilson loop
not involving ghosts can be regarded as sections of the graded
algebra bundle $\End(E)\otimes\Omega^{0,\bullet}$. Bulk observables
restricted to the Wilson loop take values in the subalgebra
$\Omega^{0,\bullet}$. We need to define the action of $Q$ on the
observables on the Wilson loop so that its restriction to the
subalgebra of bulk observables is given by (\ref{BRSTQ}). We try the
following differential operator on $E\otimes \Omega^{0,\bullet}$:
$$D={\delbar} +{\cal K}=\delbar +
\left(\begin{tabular}{cc}$\omega_{\bar I}^+d\phi^{\bar I}$ & $\Tau$\\
${\mathcal S}$ &
$\omega_{\bar I}^-d\phi^{\bar I}$\\
\end{tabular}\right),$$
where $\omega^+$ and $\omega^-$ are connection $(0,1)$-forms on
$E^+$ and $E^-$, ${\mathcal S}\in \Hom(E^+,E^-)$ and
$\Tau\in\Hom(E^-,E^+)$. The operator $D$ is a
$\bpartial$-superconnection on $E$. We will also need connection
$(1,0)$-forms on $E^+$ and $E^-$
$$\del^{\pm}=\del + \omega^{\pm}_I d\phi^I.$$

The bulk BRST operator $Q$ is not nilpotent: it squares to a gauge
transformation with a parameter $\eps^a=\kappa^{ab}\mu_{+b}$. We
should impose the same constraint on the BRST operator acting on the
observables on the Wilson loop. In particular, this means that we
need to have an action of $\frg$ on the space of smooth sections of $E$ which is compatible with the action of $\frg$ on smooth functions on $X$. The latter is given by
$$
e_a: f\mapsto V_a(f)=V^\hP\partial_\hP f,\quad e_a\in \frg,\ f\in C^\infty(X).
$$
Here hatted indices run both over holomorphic and anti-holomorphic values, e.g. $\hat
P=(P,\bar P)$.
The general form of such an action is
$$
e_a:s\mapsto \cV_a(s)=V^\hP\nabla_\hP s+ T_a s,\quad e_a\in \frg,\ s\in\Gamma(E).
$$
Here $T_a$ is an even section of $\End(E)$ which we can write as a supermatrix:
$$
T_a=\left(\begin{tabular}{cc}$t^+_a$ & 0\\ $0$ & $t^-_a$\\
\end{tabular}\right)
$$

We will require the connection $\omega^\pm$ on $E^\pm$ to be $\frg$-invariant, i.e.
$\cV_a$ should map covariantly constant sections to covariantly constant sections. This gives
\be \nabla_{\hat P}t_a^{\pm}=V_a^{\hat K}{\cal
F}^{\pm}_{\hat K \hat P}, \label{algii}\ee  Here the curvature 2-forms for $E^\pm$ are defined by
$$
{\cal F}^{\pm}=d\omega^{\pm}+\omega^{\pm}\wedge \omega^{\pm}.
$$
The requirement that the operators $\cV^a$ form a representation
of the Lie algebra $\frg$ gives \be
[t_a^{\pm},t_b^{\pm}]=f_{ab}^ct_c^{\pm}+V_a^{\hat J}V_b^{\hat K}
{\cal F}^{\pm}_{\hat J \hat K} \label{alg} \ee

We are going to impose\footnote{The idea of this construction arose
in a conversation with Lev Rozansky.} the following condition on the
Wilson-loop BRST operator $D$:
\begin{align}
D^2&=\kappa^{ab}\mu_{+a}T_b,\label{D10}\\
[D,\cV_a]&=0.\label{D20}
\end{align}
The first condition essentially says that on holomorphic sections
$D^2$ is a gauge transformation with a parameter
$\kappa^{ab}\mu_{+b}$. Indeed, for holomorphic sections $\cV_a$
reduces to
$$
\wV_a=V_a^I\nabla_I+T_a,
$$
and since $\mu_{+}\cdot\mu_+=0,$ we have
$$
\mu_+\cdot \wV=\mu_+\cdot T.
$$
Note that the differential operator $\wV_a$ is holomorphic (commutes
with $\nabla_{\bar I}$) thanks to the condition (\ref{algii}).

The second condition says that the BRST operator commutes with gauge
transformations. It is understood there that $\cV_a$ is lifted from
$E$ to $E\otimes\Omega^{0,\bullet}$. This lift is canonical: since
$\frg$ acts on $X$, it also acts on forms on $X$ via the Lie
derivative.

These conditions are equivalent to the following constraints on the
connections $\omega^\pm$ and the endomorphisms $\cS$ and $\Tau$:
\begin{align}
&\cF^\pm_{\bI\bJ}=0, &\label{D30}\\
&\nabla_{\bar I}\cS =0, &\nabla_{\bar I} \Tau =0,\\
&\cS\Tau =\kappa^{ab}\mu_{+a}t^-_b, &\Tau\cS
=\kappa^{ab}\mu_{+a}t^+_b \\
\label{dif} & V^I_a\nabla_I\Tau =\Tau t_a^--t_a^+\Tau, &
V^I_a\nabla_I{\cS} ={\cS}t_a^+-t_a^-{\cS}.
\end{align}
Here the covariant derivatives are given by
$$
\nabla_{\hP}\Tau=\partial_\hP\Tau +\omega^+_\hP \Tau-\Tau
\omega^-_\hP,\quad \nabla_{\hP}\cS=\partial_\hP\cS +\omega^-_\hP
\cS-\cS \omega^+_\hP.
$$
Note that the equations (\ref{D30}-\ref{dif}) imply, in particular,
that $E^\pm$ are holomorphic vector bundles, and $\cS$ and $\Tau$
are holomorphic bundle maps. However, we do not get a complex of
vector bundles because $\cS\Tau$ and $\Tau\cS$ are not equal to
zero, in general. This is similar to what happens in the category of
matrix factorizations arising in the Landau-Ginzburg models
\cite{LG1,LG2,LG3}.

Note also that the condition (\ref{D20}) can be simplified using
(\ref{D1}). Namely, the part of $\cV_a$ proportional to the
anti-holomorphic component of $V_a$ automatically commutes with $D$
thanks to (\ref{D30}). Therefore (\ref{D20}) is really a condition
on the holomorphic differential operator $\wV_a=V_a^I\nabla_I +
T_a$:
$$
[D,\wV_a]=0.
$$

Now let $\gamma$ be a closed curve in $M$ parameterized by $t\in
[0,1).$ The fields $\phi^{\hat I}$ specify a section of the
principal $G$-bundle $X_\cE$ with fiber $X$. Our considerations are
local, so we can choose a trivialization of $X_\cE$ and think of
$\phi^{\bar I}$ as a map from $M$ to $X$. Independence of the choice
of a trivialization will be ensured by keeping track of
gauge-invariance.

We introduce a supermanifold $\Pi {\bar T}_X$ with odd coordinates
$\eta^{\bar I}.$ There is an obvious map $\pi: \Pi {\bar T}_X
\mapsto X,$ and we may regard ${\cal K}$ as a locally-defined odd
section of the $\ZZ_2$-graded bundle $\pi^*\End(E).$ Similarly,
$\delbar$ can be interpreted as an odd vector field $\eta^{\bar
I}\partial_{\bar I}$ on $\Pi {\bar T}_X,$ and $D$ is a first-order
differential operator on $\pi^*E$. Given a map
$\Phi=(\phi,\eta):\gamma\mapsto \Pi {\bar T}_X$ and a section
$\chi_t^Idt$ of $\phi^*T_X\otimes T^*_{\gamma},$ we consider a
connection 1-form on the pull-back of $\Phi^*(E)$: \be
\label{wilsoncomplex}{\cal N}=\left(\begin{tabular}{cc}$A_t^ct_c^+
-\omega^+_{\hat I}\partial_t\phi^{\hat I} -\chi_t^N \eta^{\bar M}
{\cal F}^+_{N\bar M}$ & $-\chi_t^I\nabla_I \Tau$\\
$-\chi_t^I\nabla_I {\cal S}$ & $A_t^ct_c^- -\omega^-_{\hat
I}\partial_t\phi^{\hat I} -\chi_t^N \eta^{\bar M}
{\cal F}^-_{N\bar M}$\\
\end{tabular}\right)dt.\ee
Using (\ref{algii}) and (\ref{dif}), as well as
$D^2=\kappa^{ab}\mu_{+a}T_b$ we find
$$\delta_Q {\cal N}=-d_t({\cal K})+[{\cal N},{\cal K}].$$
Hence one can construct a BRST-invariant Wilson loop by letting
$${\mathsf W}=\STr\ {\cal U}(0,1)$$
where ${\cal U}(0,t)$ is the parallel transport operator of the
connection ${\cal N},$ i.e. the unique solution of the first order
differential equation $$(d_t -{\cal N}){\cal U}(0,t)=0$$ satisfying
${\cal U}(0,0)=\xId.$

One can also check that under a gauge transformation the connection
$\cN$ transforms as follows:
$$
\delta_\eps\cN=-[d_t-\cN,\eps^a T_a].
$$
Hence the Wilson loop is gauge-invariant.

\subsection{Wilson loops and the equivariant derived category}

In this section we reformulate the data involved in the construction
of a Wilson loop observable in purely holomorphic terms. We will see
that Wilson loops can be naturally regarded as objects of a category
which is a deformation of the equivariant derived category of
coherent sheaves on $X$. This deformation apparently is required for
the existence of a nontrivial braided monoidal structure on the
category of Wilson loops. In the case of flat $X$, the deformed
category should be equivalent to the derived category of
representations of a quantum supergroup.

Let $(A,d)=(\Omega^{0,\bullet},\bpartial)$ be the Dolbeault complex,
regarded as a $\ZZ_2$-graded differential algebra. The algebra $A$
has a Poisson bracket $\{,\}$ coming from the holomorphic symplectic
form $\Omega$. This bracket is even and $d$ is a derivation with
respect to it, i.e.
$$
d\{f_1,f_2\}=\{df_1,f_2\}\pm \{f_1,d f_2\}.
$$
The triple $(A,d,\{,\})$ is a differential Poisson algebra.

The complexification $\frg_\CC$ of the Lie algebra $\frg$ acts on
$A$ by holomorphic vector fields $W_a=V^I_a\partial_I$, which we can
regard as even derivations of $A$ commuting with $d$. They satisfy
$$
[W_a,W_b]=f^c_{ab} W_c.
$$
These derivations are Hamiltonian, in the sense that there exist
even elements $\mu_{+a}\in A$ (the moment maps) such that
$$
W_a  (f)=-\{\mu_{+a},f\}, \quad \forall f\in A.
$$
The moment maps are required to satisfy
$$
\{\mu_{+b},\mu_{+c}\}=-f^a_{bc}\mu_{+a}.
$$
It is also required that $\mu_{+a}$ is isotropic with respect to an
invariant metric $\kappa^{ab}$ on the dual vector space of
$\frg_\CC$:
$$
\kappa^{ab}\mu_{+a}\mu_{+b}=\mu_+\cdot\mu_+=0.
$$

The space of sections of the graded vector bundle $E\otimes
\Omega^{0,\bullet}$ can be regarded as a graded module $M$ over $A$.
We denote by $f\bullet m$ the action of $f\in A$ on $m\in M$. The
operator $D:M\raa M$ is an odd derivation of $M$, i.e.
$$
D(f\bullet m)=df\bullet m\pm f\bullet Dm,\quad \forall m\in M,\
\forall f\in A.
$$
The operators $\wV_a$ on the space of smooth sections of $E$ make $M$ into a
$\frg_\CC$-equivariant $A$-module. That is, $M$ is an
$\frg_\CC$-module, and this module structure is compatible with the
$\frg_\CC$-module structure on $A$ in the following sense:
$$
\wV_a(f\bullet m)=W_a(f)\bullet m+f \bullet \wV_a(m),\quad \forall
m\in M,\ \forall f\in A.
$$
Finally, $D$ satisfies the following two conditions:
\begin{align}
[D,\wV_a]&=0,\label{D1}\\
D^2 &=\kappa^{ab} \mu_{+a} \wV_b.\label{D2}
\end{align}
If we replace (\ref{D2}) with a condition $D^2=0$, then the triple
$(M,D,\wV_a)$ becomes an equivariant differential graded module over
the supercommutative DG-algebra $(A,d,W_a)$ with a
$\frg_\CC$-action. Note that the Poisson bracket, the moment maps,
and an invariant metric on $\frg_\CC^*$ are not needed to define
such a module. But these data are needed to define the deformed
category whose objects label the Wilson loops in the CSRW model.

Let us say a few words about morphisms in the deformed category.
From the physical point of view these are observables inserted at
the joining point of two Wilson lines labeled by objects
$(M_1,D_1,\wV_{1a})$ and $(M_2,D_2,\wV_{2a})$. A morphism is
therefore an $A$-module morphism $\phi: M_1\raa M_2$. This space has
a natural $\ZZ_2$ grading. BRST operator $Q$ defines an odd
derivation $D_{12}$ on the space of morphisms:
$$
D_{12}\phi=D_2\circ\phi\pm \phi\circ D_1.
$$
This derivation is not nilpotent, rather
$$
D_{12}^2\phi=\kappa^{ab}\mu_{+a}\left(\wV_{2b}\circ\phi-\phi\circ\wV_{1b}\right).
$$
It is nilpotent on the subspace of equivariant morphisms, i.e. those
$\phi$ for which the expression in parentheses on the r.h.s. of the
above equation identically vanishes. Thus it is possible to define a
differential-graded category whose objects are as above, morphisms
are equivariant morphisms of $A$-modules, and the differential on
the space of morphisms is $D_{12}$.

From the physical viewpoint this prescription for computing the
space of observables is not quite correct since it does not include
the Faddeev-Popov ghosts. The correct prescription is to tensor the
space of morphisms with $\bigwedge^\bullet \frg_\CC^*$ (the algebra
of ghost fields) and compute the cohomology of the operator $\hQ$.
It has the form
$$
\hQ=D_{12}+Q_{CE}+\mu_{+a}\kappa^{ab}\frac{\partial}{\partial c^b},
$$
where $Q_{CE}$ is the Chevalley-Eilenberg differential corresponding
to the $\frg_\CC$-module $\Hom_A(M_1,M_2)$. One can check that
$\hQ^2=0$.

Replacing $Q$ with $\hQ$ should be thought of as passing to the
derived version of the category. Indeed, if we formally set
$\kappa^{ab}=0$, $(M,D,\wV_a)$ becomes the usual equivariant
DG-module over $(A,d,W_a)$, and tensoring $\Hom_A(M_1,M_2)$ with the
ghost algebra and adding $Q_{CE}$ to the differential $D_{12}$ is
the standard way to get a free resolution of the $\frg_\CC$-module
$\Hom_A(M_1,M_2)$.

Let us make a few comments about the deformed category. First of
all, since $D$ is an odd derivation of $M$, $D^2$ is an even
endomorphism of the module $M$. Although $\wV_a$ is not an
endomorphisms of $M$, the combination $\kappa^{ab}\mu_{+a}\wV_b$ is,
thanks to the condition $\mu_+\cdot\mu_+=0.$ Indeed, if $f$ is an
arbitrary element of $A$, then
$$
[\kappa^{ab}\mu_{+a}\wV_b,f]=\kappa^{ab}\mu_{+a}[\wV_b,f]=\kappa^{ab}\mu_{+a}
W_b(f)=-\kappa^{ab}\mu_{+a}\{\mu_{+b},f\}=-\frac12\{\mu_+\cdot\mu_+,f\}=0.
$$

Second, the two conditions (\ref{D1},\ref{D2}) are compatible:
\begin{multline}
0=[D^2,\wV_c]=[\kappa^{ab}\mu_{+a}\wV_b,
\wV_c]=\kappa^{ab}\left(\mu_{+a}[\wV_b,\wV_c]-[\wV_c,\mu_a]\wV_b\right)=\\
=\kappa^{ab}\left(f^d_{bc}\mu_{+a}\wV_d+\{\mu_{+c},\mu_{+a}\}\wV_b\right)=\left(\kappa^{ab}f^d_{bc}+
\kappa^{db}f^a_{bc}\right)\mu_{+a}\wV_d=0.
\end{multline}
In the second line we used the $\frg$-invariance of $\kappa^{ab}$.

Third, the fact that the Poisson bracket came from a symplectic form
was not important. Thus the deformation of the equivariant derived
category makes sense in the context of Poisson manifolds.
Furthermore, the condition $\mu_+\cdot\mu_+=0$ can be relaxed to the
condition that $\mu_+\cdot\mu_+$ have vanishing Poisson brackets
with any element of $A$. In the case when the Poisson bracket comes
from a symplectic form this generalization is not very significant,
since the Poisson center of $A$ consists of constants. It becomes
interesting when the Poisson bracket is degenerate.

Fourth, there is a $\ZZ$-graded version of the story. If $A$ and $M$
are $\ZZ$-graded and $d:A\raa A$ and $D:M\raa M$ have degree $1$,
then for the condition (\ref{D2}) to make sense $\mu_{+a}$ must have
degree $2$. Since $W_a$ has degree $0$, the Poisson bracket must
have degree $-2$. Such a situation is realized, for example, when
$A$ is the Dolbeault complex of the cotangent bundle of a complex
manifold $Y$, provided that we put the linear coordinates on the
fibers in degree $2$. Above we dealt with a special case of this,
namely $Y=G_\CC/B$.

Fifth, while the algebra $A$ occurring in the CSRW model is the
algebra of forms on a smooth complex manifold, the definition of the
deformed equivariant derived category given above is purely
algebraic and makes sense in greater generality.

Sixth, the deformation which takes us from the equivariant derived
category of $(A,d,W_a)$ to the category defined above should be
thought of as a quantum deformation. Indeed, absorbing the
Chern-Simons level $k$ into the metric $\kappa_{ab}$ we see that
$\kappa^{ab}$ is proportional to the Planck constant $1/k$ of the
CSRW model. The classical limit is the limit $\kappa^{ab}\raa 0$, in
which case the category of Wilson loops reduces to the usual
equivariant derived category.

\subsection{Examples of Wilson loop obsevables}
For flat $X$ the CSRW model is equivalent to a super-Chern-Simons
theory. Therefore there is a Wilson loop operator for every
finite-dimensional representation $R$ of the Lie superalgebra
$\frG$:
$${\cal W}_R=\STr_R~Pe^{\oint A^aM^{(R)}_a+\chi^I\lambda^{(R)}_I}.$$
Here $M^{(R)}_a$ and $\lambda^{(R)}_I$ are endomorphisms of a graded
vector space $R$ representing bosonic and fermionic generators of
$\frG$. This is a special case of our general construction in
Section 5.1. To see this, we take $X$ to be the odd part of $\frG$.
The even part $\frg$ of $\frG$ acts on $X$ linearly. We take $E$ to
be a trivial vector bundle over $X$ with fiber $R$. Since the origin
of $X$ is invariant under $\frg$, we can specify an action of $\frg$
on $E$ by specifying its action on the fiber of $E$ over the origin.
We take the tautological action of $\frg$ on $R$ ($R$ is a
representation of $\frG$ and therefore a representation of its even
subalgebra $\frg$). Finally we let
$$
D=\bpartial -\phi^I \lambda^{(R)}_I.
$$
It is not entirely clear if for flat $X$ any object in the category
of Wilson loops is isomorphic to an object of this special form. We
conjecture this to be the case.

For a curved target space it is rather difficult to find nontrivial
examples of BRST-invariant Wilson loops. One could start with
equivariant holomorphic vector bundles or complexes of vector
bundles and try to deform them. In the special case $X=T^*\CC\PP^1$
we will now exhibit a family of Wilson loops for which deformation
is unnecessary. Let us take $E$ to be a holomorphic line bundle over
$X$ with a $SL(2,\CC)$-action and let $D$ be the usual
$\bpartial$-connection on $E$ (i.e. we let $\Tau=\cS=0$). If we want
$E$ to be an object of the category of Wilson loops, then the
endomorphisms $T_a$ must satisfy
\begin{equation}\label{line}
\kappa^{ab}\mu_{+a} T_b=0.
\end{equation}
They must also satisfy
\be \label{lineii}
f_{ab}^cT_c+V_a^{\hat J}V_b^{\hat K} {\cal F}_{\hat J \hat
K}=0,\quad \del_{\hat P}T_a=V_a^{\hat K}{\cal F}_{\hat K \hat P}.\ee

Since $T^*\CC\PP^1$ is simply-connected, we can specify $E$ together
with an $SU(2)$-invariant connection by specifying an
$SU(2)$-invariant $(1,1)$ form $\cF$ whose periods are integral
multiples of $2\pi$. We take
$$
\cF=(-i) n {\hat J},\quad n\in\ZZ
$$
where ${\hat J}$ is a K\"ahler form on $T^*\CC\PP^1$ normalized so
that the integral of $\hat J$ over the zero section of $T^*\CC\PP^1$
is $2\pi$. The corresponding line bundle $E=\cL^n$ on $T^*\CC\PP^1$
restricts to $\cO(n)$ on $\CC\PP^1$. Given this $\cF$ the
endomorphisms $T_a$ are uniquely determined:
$$
T_a=(-i) n {\hat \mu}_{3a},
$$
where ${\hat\mu}_{3a}$ is the moment map for $\hat J$. The condition
$\mu_{+}\cdot T=0$ is satisfied thanks to (\ref{moments}). Note that
at the point $z=b=0$ we have
$$
T_1=T_2=0,\quad T_3=(-i)n.
$$
That is, the point $z=b=0$ is fixed by a $U(1)$ subgroup of $SU(2)$,
and the fiber of the line bundle $\cL^n$ over this point transforms
in the representation with charge $n$. By $SU(2)$ symmetry this is
true for any other point on the zero section of $T^*\CC\PP^1$: the
fiber of $\cL^n$ over a point transforms in a charge $n$
representation of the $U(1)$ subgroup preserving this point.

Let us call $\sfW_n$ the Wilson loop corresponding to the line
bundle $\cL^n$. At the classical level it is clear that the product
of $\sfW_n$ and $\sfW_m$ is $\sfW_{n+m}$. There may be no quantum
corrections to this result, since the Wilson loop operator $\sfW_n$
cannot be deformed. This follows, for example, from the fact that
the endomorphisms of $\sfW_n$ regarded as an object of the
equivariant derived category of $T^*\CC\PP^1$ are the same as the
endomorphisms of the trivial line bundle. This implies that there
are no endomorphisms of $\sfW_n$ with ghost number one whose
descendants could be used to construct an infinitesimal deformation
of $\sfW_n$.

So far, the Chern-Simons level $k$ did not play any role in the
discussion. One place where it shows up is in the braiding
properties of the Wilson loops. The braiding phase is computed by
taking Wilson loops $\sfW_{n_1}$ and $\sfW_{n_2}$ along the closed
curves $\gamma_1$ and $\gamma_2$ in $R^3$ with linking number one
and computing the correlator $\langle \sfW_{n_1} \sfW_{n_2}\rangle.$
Using the well-known property of the Green-function $G_{\mu \nu}$
\cite{RozanskyWitten} we find\footnote{$\langle A^a_{\mu}(x_1)
A^b_{\nu}(x_2)\rangle=-\hbar G_{\mu \nu}(x_1,x_2) \kappa^{ab}, \quad
\oint_{\gamma_1}dx_1^{\mu}\oint_{\gamma_2}dx_2^{\nu}G_{\mu
\nu}(x_1,x_2)=1.$} at leading order in $1/k$ expansion the phase
$e^{\pi i n_1 n_2\over k}$ plus BRST-exact terms. The phase is
essentially the same as in ordinary Chern-Simons perturbation theory
for Wilson loops, with generators of the Lie algebra in a particular
representation replaced with the endomorphisms $T_a$. We will see
below that there can be no corrections to this phase at higher order
in perturbation theory.

Another finite-$k$ effect is a periodic identification among Wilson
loops: $\sfW_n$ is isomorphic to $\sfW_{n+2k}$ in the category of
Wilson loops. To see this, it is sufficient to exhibit an invertible
morphism between $\sfW_{2k}$ and the trivial Wilson loop
corresponding to a trivial line bundle on $T^*\CC\PP^1$. The space
of such morphisms can be thought of as the space of local
observables which can be inserted at the endpoint of $\sfW_{2k}$
(which is now a Wilson line rather than a Wilson loop).
Equivalently, by state-operator correspondence in the CSRW theory,
it is the space of states of the theory on the space-time of the
form $S^2\times\RR$, with a Wilson line $\sfW_{2k}$ inserted at
$\{p\}\times \RR$ where $p$ is point on $S^2$.

The quantization is simplified by the fact that all fermionic fields
as well as the bosonic field $b$ parameterizing the fiber direction
in $T^*\CC\PP^1$ are massive, and the only matter field zero mode is
that of the field $z$ which parameterizes the zero section of
$T^*\CC\PP^1$. Further, with all massive fields in their ground
state, the Gauss law constraint in the presence of a Wilson loop
reads
$$
\frac{k}{2\pi} \kappa_{ab}F^b=iT_a\delta^2(p)=n{\hat \mu}_{3~a}(z)\delta^2(p),
$$
where $F$ is the curvature of the $G$-connection $A$. Thus the gauge
field is also determined by $z$. We also see that for $n\neq 0$
there is a magnetic flux on $S^2$ proportional to $n$. More
precisely, for a fixed $z$ the gauge group is broken down to a
$U(1)$ (the stabilizer of $z$), and the Gauss law constraint says
that the gauge bundle reduces to this $U(1)$ subgroup and its first
Chern class is $n/2k$. Clearly, this makes sense only if $n$ is an
integer multiple of $2k$, $n=2km$.

The conclusion is that we need to quantize $\CC\PP^1$ parameterized
by the zero mode of $z$. The symplectic form arises both from the
Chern-Simons part of the action and the Wilson line. It is clear
that it will be an integer multiple of the curvature 2-form of the
line bundle $\cO(1)$. If the coefficient is $l$, then quantization
gives $l+1$-dimensional space of states which furnishes an
irreducible representation of $SU(2)$. On the other hand, Gauss law
constraint guarantees that this space will be an $SU(2)$-singlet.
This means that the Chern-Simons contribution must exactly cancel
the contribution from the Wilson loop, so that $l=0$. (One can
verify this explicitly). This proves that there is a unique monopole
operator on the endpoint of any Wilson loop $\sfW_n$ such that $n$
is divisible by $2k$. The invertibility of this morphisms then
follows from general axioms of 3d TFT. Alternatively, it is clear
that when an open Wilson line terminated by monopoles shrinks to a
point, the monopoles just annihilate each other and one is left with
the identity operator in the bulk theory.

Note that the phase arising from braiding must be invariant under
$n\mapsto n+2k$. This rules out any higher-order perturbative
corrections to the phase computed above.

\section{Concluding remarks}

In the previous section we have determined the category of Wilson
loop operators in the general CSRW model and provided some examples
of Wilson loops. On general grounds, we expect this category to be a
braided monoidal category. It would be very interesting to compute
the braiding, which is bound to be nontrivial because of
Chern-Simons terms. The most interesting case is that of
$X=T^*(G_\CC/B)$, since this theory is similar in many respects to
the ordinary Chern-Simons theory. In particular, it has no
nontrivial local operators and its partition function is finite and
provides new invariants of three-manifolds. It would be very
interesting to study the structure of the perturbation series in
this and other CSRW models, since it provides new solutions of the
IHX relation \cite{IHX}. One might also speculate that the knot
invariants arising from the Wilson loop correlators in the CSRW
model are related to quantum group knot invariants at non-primitive
roots of unity.

\section{Appendix A}
Here we explain how to derive the BRST transformations (\ref{BRSTQ})
of the CSRW model by twisting supersymmetry transformations in $N=4$
$D=3$ superconformal theory constructed by Gaiotto and Witten. For
simplicity we do this for a flat target $X$. As we explained in
section 2, BRST transformations (\ref{BRSTQ}) also work for a curved
target with appropriate moment maps.

Let $\theta_A^{\dot B ~\alpha}$ be a parameter of supersymmetry
transformation, where $A({\dot B})$ runs over a doublet of
$SU(2)_R(SU(2)_N)$, and $\alpha$ runs over a doublet of the Lorentz
symmetry $SU(2)_L.$ The supersymmetry transformations of the
Gaiotto-Witten model are \be\label{susy} \delta Q_A^{\bar
I}=\theta_A^{\dot B ~\alpha}\lambda^{\bar I}_{\dot B ~\alpha}\ee
$$\delta \lambda^{\bar I}_{\dot A \alpha}=\theta_{\dot A}^{B~ \beta}\sigma^{\mu}_{\alpha \beta}D_{\mu}Q^{\bar I}_B
+{1\over 3}\theta^B_{\dot A ~\alpha}V_a^{{\bar
I}~C}\mu_{b~CB}\kappa^{ab}$$
$$\delta (A^a_{\mu})\sigma^{\mu}_{\alpha \beta}=\kappa^{ab}
\theta^{A\dot B}_{(\alpha}\lambda^{\bar I}_{\beta)\dot B}
\Omega_{\bar I \bar J}V_{b~ A}^{\bar J}.$$ Here
$$Q_1^{\bar I}=\phi^{\bar I},\quad Q_2^{\bar I}=2\Omega^{\bar I \bar J}g_{\bar J K}\phi^{K},\quad
V_1^{a~\bar I}=V^{a~\bar I},\quad V_2^{a~\bar I}=2\Omega^{\bar I
\bar J}g_{\bar J K}V^{a~K},$$
$$\mu_{a~11}=2\mu_{a~-},\quad \mu_{a~12}=-2i\mu_{a~3},\quad
\mu_{a~22}=2\mu_{a~+}.$$ We work in conventions
$\Omega=\half \Omega_{IJ}d\phi^I\wedge d\phi^J,\quad J=ig_{I \bar J}d\phi^I\wedge d\phi^{\bar J}$
so that hyper-K\"ahler structure implies $\Omega^{\bar I \bar J}=-{1\over 4} g^{\bar I K}\Omega_{KL}g^{L \bar J}$ and for flat $X$ we have $g_{I \bar J}=\half \delta_{I \bar J}.$

We twist by identifying $SU(2)_L$ and $SU(2)_N$. Then
$$\theta_{1}^{\dot A ~\beta}=\theta^{2~\dot A ~\beta}=-\theta_{BRST} \epsilon^{\dot A ~\beta}.$$
We may set all other supersymmetry variation parameters to zero,
keeping only the BRST parameter $\theta_{BRST}.$ Then we write
fermions in terms of the fields of the twisted model as \be
\label{susyii}\lambda^{\bar I}_{\dot A ~\alpha}=\epsilon_{\dot A
~\alpha}\eta^{\bar I}+ 2\sigma^{\mu}_{\dot A ~ \alpha}\Omega^{\bar I
\bar J}g_{\bar J K}\chi_{\mu}^{K}\ee and plug (\ref{susyii}) into
(\ref{susy}) to obtain BRST transformation (\ref{BRSTQ}).

\end{document}